# In-situ Raman spectroscopy of the graphene/water interface of a solution-gated field effect transistor: electron-phonon coupling and spectroelectrochemistry

J. Binder[1*], J. M. Urban[1†], R. Stepniewski[1], W. Strupinski[2] and A. Wysmolek[1]

[1] Faculty of Physics, University of Warsaw, Pasteura 5, 02-093 Warsaw, Poland
[2] Institute of Electronic Materials Technology, Wólczyńska 133, 01-919 Warsaw, Poland

**Abstract**

We present a novel measurement approach which combines the electrical characterization of solution-gated field effect transistors based on epitaxial bilayer graphene on 4H-SiC (0001) with simultaneous Raman spectroscopy. By changing the gate voltage, we observed Raman signatures related to the resonant electron-phonon coupling. An analysis of these Raman bands enabled the extraction of the geometrical capacitance of the system and an accurate calculation of the Fermi levels for bilayer graphene. An intentional application of higher gate voltages allowed us to trigger electrochemical reactions, which we followed in-situ by Raman spectroscopy. The reactions showed a partially reversible character, as indicated by an emergence / disappearance of peaks assigned to C-H and Si-H vibration modes as well as an increase / decrease of the defect-related Raman D band intensity. Our setup provides a highly interesting platform for future spectroelectrochemical research on electrically induced sorption processes of graphene on the micrometer scale.

## 1. Introduction

Graphene is considered a very promising candidate for a plethora of applications.[1] The use of graphene as active material in sensing applications was recognized early and led to many interesting approaches.[2-4] Solution-gated graphene field effect transistors (SGFET) are one of them.[5-10] Here, the solution can be seen to fulfill a double task. Firstly, it plays a role equivalent to that of regular thin film dielectrics in common transistors. In contrast to these conventional dielectrics, an electrical

---

[*] Corresponding author. E-mail: Johannes.Binder@fuw.edu.pl,
[†] currently at: Institute of Physical Chemistry, University of Freiburg, Albertstraße 23a, 79104 Freiburg, Germany

double layer forms at the graphene/solution interface.[11] A high potential drop across this thin double layer leads to a strong gating effect, which allows the control of the carrier concentration in a broad range. Secondly, for sensor applications, the solution provides a carrier medium for the substance to be detected. For a SGFET, the detection is realized by recording a change in the resistance of graphene in direct contact with the solution. The processes leading to the underlying change in carrier concentration depend on the quality and surface characteristics of graphene[12] and can be tailored by specific functionalization of the carbon surface.[13-15]

Raman spectroscopy is known to be a versatile tool for the characterization of carbon allotropes[16] and has become a workhorse technique in the graphene research community.[17,18] This technique allows to draw conclusions about the number of carbon layers [17-19], the stacking order [20,21], strain [22-24] and carrier concentration.[25-29]

Spectroelectrochemistry is a combination of one or more spectroscopy techniques (e.g. optical absorption, Raman spectroscopy, electron paramagnetic resonance) and reaction-oriented electrochemistry.[30-32] The reduction of graphene oxide has been investigated in-situ using an aqueous electrolyte and Raman spectroscopy by focusing the laser beam on the sample through the electrolyte[33] or by using special spectroelectrochemical cells.[34] In-situ Raman studies have also been performed for other carbon allotropes, e.g. on the hydrogenation of nanoporous carbon.[35] A study of a bilayer graphene FET on $SiO_2$ showed that the Fermi level position of graphene strongly impacts the chemisorption processes of oxygen.[36] Another work proved that the etching of graphene edges upon UV irradiation depends strongly on the gate voltage applied.[37] These works highlight the special role that graphene can play in chemisorption processes with the Fermi level as a switch for specific reactions.

In this work, we present results obtained by a novel measurement approach. We were able to carry out simultaneous measurements of the SGFET electrical characteristics and Raman spectroscopy of graphene on a 4H SiC(0001) substrate in an aqueous 100 mM NaCl solution. This solution was used as the liquid environment in order to facilitate the comparison with previous SGFET reports [14,38] as well as to relate the reported results to biologically relevant conditions. By taking advantage of the fact that the SiC substrate is transparent for the laser wavelength used in this work (532 nm), we could perform Raman spectroscopy measurements of graphene in contact with the solution through the SiC substrate. This measurement scheme with the laser exposure from the back of the sample has been used by Fromm et al.[39] for a Raman study of an epitaxial graphene transistor with a conventional dielectric. In our case, the epitaxial graphene on SiC acted as a semitransparent working electrode. Our setup allows the electrochemical reactions to be followed by monitoring the evolution of the Raman signal, which builds a powerful tool for investigations in spectroelectrochemistry.

The aim of this work is to present the fundamental characterization of the above described setup including an analysis of the electron-phonon coupling in bilayer graphene. The Raman features resulting from the resonant electron-phonon coupling allowed us to deduce the actual Fermi level shift in graphene. We could therefore recalibrate our gate voltages to Fermi level shifts without the need to use parallel plate capacitor models based on estimations of the dielectric constants of the solution close to the surface. We also compared two kinds of graphene bilayer samples: a hydrogen-intercalated and a non-intercalated sample. Proof-of-principle spectroelectrochemical measurements of the chemisorption of hydrogen on epitaxial graphene are presented, illustrating the capabilities of our approach. The obtained results are of high interest for the selective functionalization of graphene as well as for hydrogen storage considerations.

2. Experimental

A horizontal hot wall reactor was used for the graphene growth at a temperature of about 1600ºC. The 1 x 1cm 4H-SiC (0001) substrates (semi-insulating, on axis) were annealed in $H_2$ atmosphere in order to clean the surface prior to growth. Si sublimation from the surface was blocked with the use of specific dynamic argon flow conditions, as described in ref 40. For this CVD process on SiC, propane was used as the carbon precursor. Hydrogen intercalation was achieved by annealing in $H_2$ after the growth to form quasi-free standing graphene.[41-43]

The final transistor structure (hall bar geometry) was fabricated using a three-layer photolithography process (inset Figure 1). In the first step, an Au/Pd alloy (1:1) was deposited (about 30 nm) using sputter deposition and a consecutive lift-off. The second process step involved the etching of excess graphene using Ar-ion milling, for which the lithography resist served as the etch mask. In order to ensure that the metal contacts were not in contact with the solution, a third step had to be employed. Here, a negative epoxy-based resist (SU8) was used to cover the whole sample; only the active area and the contact pads, which are outside of the solution basin, were left uncovered. Electrical interconnections to the sample contact pads, which are not in contact with the solution (see Figure 1), were established by Ag silver paste. The O-ring/Pt/glass sample holder, together with the micro-structured graphene sample built our SGFET.

The experimental setup that enables the combined Raman and electrical measurements is depicted in Figure 1. In this configuration, the graphene sample was placed upside down on a fluoroelastomer O-ring. The O-ring was mounted on a glass microscope slide covered with an 8 nm thick layer of Pt (sputter deposition). The thin layer of Pt formed a semitransparent electrode, which allowed to minimize the impact of reflected laser light on the Raman measurements. This semitransparent electrode was then divided into two areas using a diamond scribe. As a result, the two regions were electrically isolated from each other and acted as the counter and quasi-reference electrode. Hence, we used a three-electrode setup which allows the separation of the voltage measurement from the current flow facilitating stable measurement conditions.[44,45] All the results presented in this work were obtained with a 100 mM NaCl / deionized water (Millipore) solution as electrolyte.

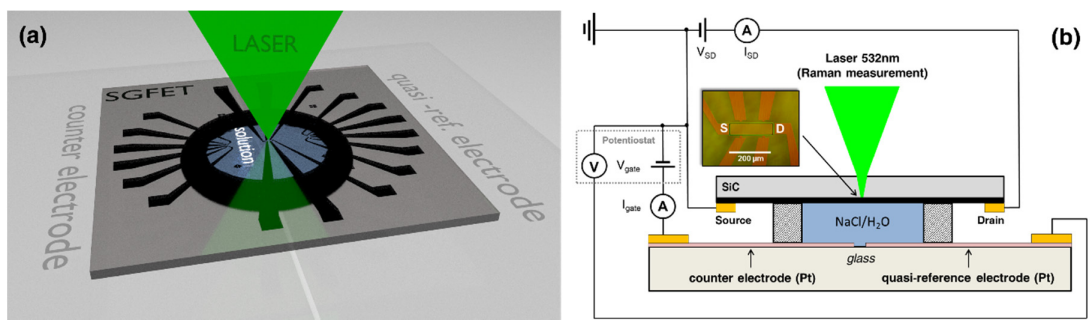

**Figure 1.** Schematic drawings of the experimental setup: (a) 3D visualization, (b) cross section with the electrical circuitry. The measurements were performed using a three-electrode setup with a Pt counter electrode and a Pt quasi-reference electrode. The inset in (b) shows a microscope image of the transistor structure.

The Raman measurements were performed using a Horiba Jobin Yvon T64000 confocal spectrometer in backscattering geometry. The 2nd harmonic of a Nd:YAG laser was used as excitation source (532 nm). The laser power was kept below 5mW and a 50x microscope objective (NA: 0.5) was employed for all the measurements reported. The electrical measurements were performed with two source measure units (Agilent B2901A, Keithley 2400).

## 3. Results and discussion
### 3.1. Electrical characteristics and simultaneous Raman spectroscopy

The electrical characteristics of a SGFET based on a non-intercalated bilayer graphene sample are shown in Figure 2. One can clearly see an ambipolar behavior typical for gated graphene devices. The charge neutrality point (CNP) for this sample was located at a gate voltage of around +0.18 V. It should be noted that the voltage notation throughout this report follows the standard convention for solid state transistor measurements. This means that the gate voltage was applied between the gate electrode and the source, while for electrochemical measurements the convention is opposite, referring voltages to the reference electrode. The shift of the CNP towards positive gate voltage values can be explained by the differences of the work functions of graphene and the Pt reference as well as the doping level of graphene. Epitaxial graphene on SiC (0001) is normally n-type doped,[40] but contact with water gives rise to an additional chemical p-type doping.[46] Furthermore, the CNP position can be strongly influenced by the presence of ions in the solution.[47,48] The leakage current across the solution was smaller than 10 nA throughout the measurement.

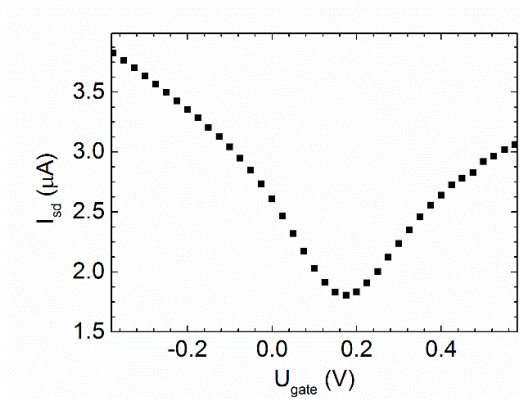

**Figure 2.** Source-drain current $I_{sd}$ for a constant source drain voltage ($U_{sd}$ = 50mV) as a function of the applied gate voltage $U_{gate}$.

Our method enabled us to collect a set of data from two independent measurement techniques at the same time and under exactly the same conditions. The results shown in Figure 2 were obtained simultaneously with Raman measurements presented in Figure 3. The Raman spectra shown were corrected for the SiC background (see Supplementary Data for more details). As can be seen in Figure 3 (a), the Raman spectra show a clear shift of G band energies. It is commonly accepted that these shifts are correlated with changes in the electron-phonon coupling strength, caused by the alternation of the Fermi level induced by the applied gate voltage.[25-28] The observed shape and the width of the 2D band clearly indicated bilayer graphene (S1, Supplementary Data).

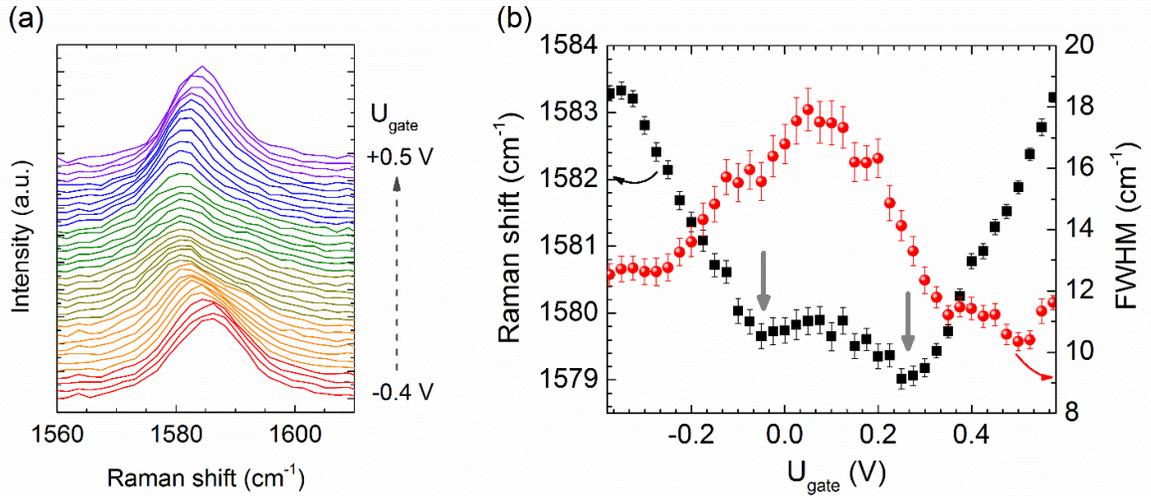

**Figure 3**. (a) Raman spectra of the G band region as a function of the applied gate voltage. The spectra were shifted for clarity and the SiC background has been subtracted. (b) Results of Lorentzian fits to the G band shown in (a); *black squares:* G band Raman shift, *red circles:* FWHM of the G band

The results of the G band analysis (single Lorentzian fit) for this measurement are shown in Figure 3 (b). A softening of the phonon can be clearly observed (*black squares*), which stems from the fact that the electron-phonon coupling is most efficient when the Fermi level is exactly at or close to the Dirac point.[25-28] The broadening of the G band (*red circles*) is due to Landau damping and results in peaks that are broadened when the Fermi energy is smaller than half the G phonon energy.[27] Interestingly, two additional minima of the phonon energy can be observed (arrows in Figure 3 (b)). We ascribe these minima to the case when the Fermi level is tuned to half the G band phonon energy and a resonance of the electrical transitions with the G band phonon is obtained.[49, 50] The minima are expected to be more pronounced at low temperatures and give rise to a logarithmic-like decrease in phonon energy.[51] Since our measurements were performed at room temperature, these features are smeared out, but nevertheless can still be clearly identified. The striking observation of these resonance-related features for epitaxial graphene even at room temperature is most likely enabled by the fact that we use bilayer graphene. The electronic bands of bilayer graphene are parabolic-like close to the Dirac point. Therefore, an inhomogeneity in carrier concentration results in a smaller broadening of the Fermi level for bilayer than for monolayer graphene, which makes the resonance effect more robust.[49]

The CNP in Figure 3 (which is situated at the half distance between the two minima) is slightly shifted as compared to the electrical measurements in Figure 2. This can be understood by taking into consideration the differences of the probed area for both experiments. In the case of the optical measurement a laser with a spot size of about 1 μm was used whereas the whole transistor area contributed to the electrical response. Such variations of the local initial doping can be explained by the interaction of graphene with the stepped terrace structure of the SiC substrate. [29,52]

### 3.2. Geometrical capacitance extraction

In order to convert the applied gate voltages to Fermi level positions ε, we had to account for the density of states (DOS) $D(\varepsilon)$ of bilayer graphene. An expression for bilayer graphene can be derived theoretically, as described by Ando[50]

$$D(\varepsilon) = \frac{2}{\pi \gamma^2} \left[ \frac{\Delta}{2} + \varepsilon + \theta(\varepsilon - \Delta)\left(\varepsilon - \frac{\Delta}{2}\right) \right] \quad (1)$$

where $\gamma = (\sqrt{3}/2)a\gamma_0$, $\Delta = \gamma_1$. Here, $a = 2.46$ Å is the lattice constant, $\theta(x)$ is the step function and $\gamma_0 \approx 3.16\ eV$, $\gamma_1 \approx 0.39\ eV$ are the tight-binding parameters describing the nearest-neighbor hopping and the interlayer coupling. For further analysis, we restrict ourselves to Fermi energies $\varepsilon_F < \gamma_1$, which means that we only take the lower energy bands into account. Bearing in mind the interpretation of the minima observed in Figure 3 (b), this approximation is reasonable. These minima should appear at half the G band phonon energy, which is about 98 meV and thus in that regime, $\varepsilon_F < \gamma_1$ is satisfied. With $\alpha = \frac{1}{\pi \gamma^2}$ this yields

$$D(\varepsilon) = \alpha\Delta + 2\alpha\varepsilon \quad (2)$$
$$n(\varepsilon) = \alpha\Delta\varepsilon + \alpha\varepsilon^2 \quad (3)$$

where $n(\varepsilon)$ is the carrier concentration.[50]

In order to assess the electrochemical gating effect quantitatively, one has to include the geometrical capacitance as well as the quantum capacitance of graphene in the calculations.[7,27,45,53,54] The quantum capacitance of graphene reflects the influence of the DOS on the capacitance of the system. To account for that, one has to use a model of two capacitors connected in series. For common back-gated graphene transistors, the geometrical capacitance is much smaller than the quantum capacitance and the latter can be neglected. However, electrochemical gating gives rise to very high geometrical capacitances that are of the same order of magnitude as the quantum capacitance. Thus, both capacitance types had to be taken into account.

Since it is our goal to convert the applied gate voltages into a Fermi level shift (or carrier concentration), we use the DOS of bilayer graphene $D(\varepsilon)$[50] to calculate the quantum capacitance. With the use of $C_q = e^2 D(\varepsilon)$ as a definition for the quantum capacitance[53,54], we obtain

$$U_{gate} = \frac{n(\varepsilon)e}{C_g} + \frac{n(\varepsilon)e}{C_q} = \frac{n(\varepsilon)e}{C_g} + \frac{n(\varepsilon)}{eD(\varepsilon)} \tag{4}$$

where $n(\varepsilon)$ is the carrier concentration and $C_g$ the geometrical capacitance.

This allows us to extract the geometrical capacitance of our system. To achieve this, we used the two characteristic points (minima indicated by arrows in Figure 3 (b)), which are the signatures of the resonant electron-phonon coupling. The gate voltage difference between these two points corresponds to a change of ε equal to the energy of the G phonon. By inserting this experimentally determined value into (4), we obtain a geometrical capacitance of $C_g$= 6,65 µF/cm². (See S5 in the Supplementary Data for more details). This result is in accordance with the values of double layer capacitance previously reported for SGFET systems[55].

### 3.3. Carrier concentrations

Using the geometrical capacitance determined in the previous section, we could calculate the Fermi level shifts using equations (2), (3) and (4). This approach yields a result that does not rely on any assumptions concerning the effective dielectric constant of the aqueous solution in the vicinity of the graphene surface. Hence, our method allows to avoid the uncertainties related to the conversion of gate voltages into actual Fermi levels, whenever the features related to the resonant electron-phonon coupling can be observed. Figure 4 shows the Raman G band energy and FWHM as a function of the Fermi level.

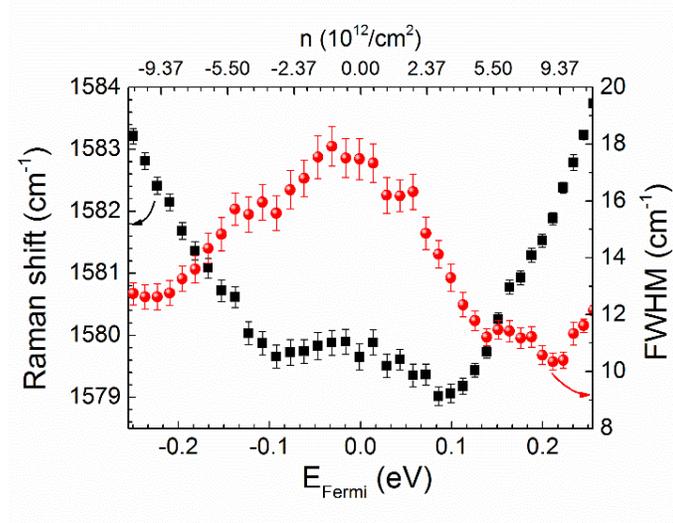

**Figure 4.** G band Raman shift (*black squares*) and FWHM of the Raman G band (*red circles)* as a function of Fermi energy and carrier concentration.

An important asset of our experiment is that we carry out simultaneous electrical and optical measurements. Thus, we are able not only to calculate the carrier concentration for the Raman spectroscopy measurements as described above, but we can also transfer the already obtained calibration to the electrical measurements, since they were performed under the exactly same conditions. Figure 5 (a) presents the calibrated electrical characteristics.

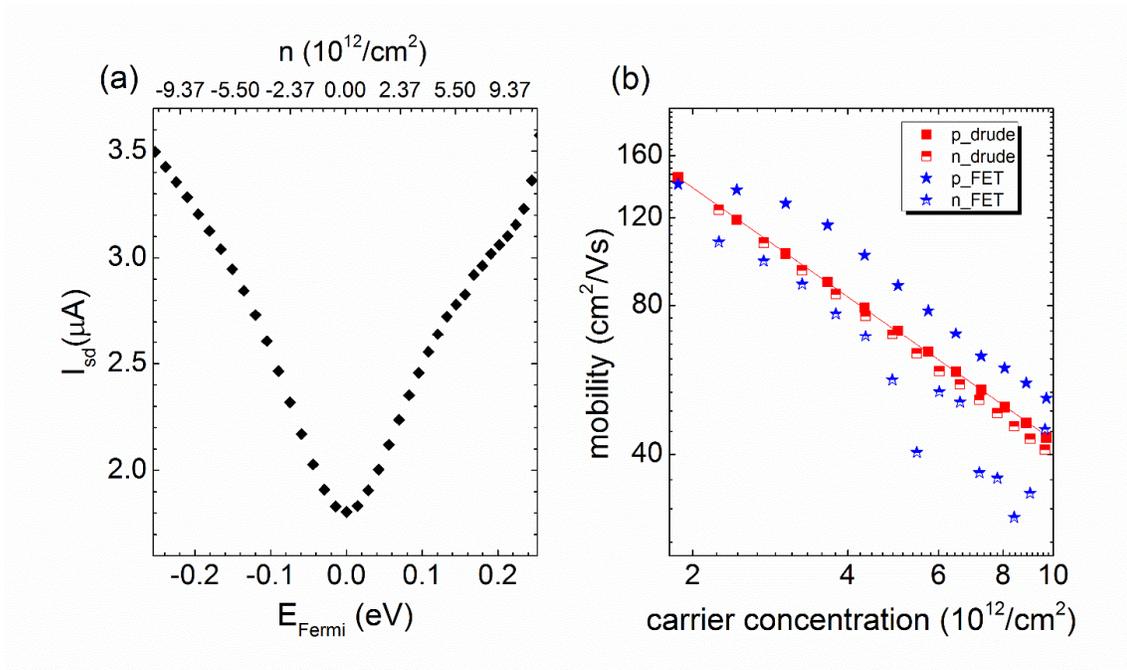

**Figure 5.** (a) Source-drain current $I_{sd}$ ($U_{sd}$ = 50mV) as a function of the Fermi energy and the carrier concentration after calibration. (b) Mobility as a function of carrier concentration; *red squares:* Mobilities derived using the Drude model; *blue stars:* Field effect mobilities; *red line:* power law fit $\mu \propto n^{-0.72}$

With the electrical characteristics at hand, we can draw conclusions about the carrier mobility of our device. We employ two models for the estimation of the carrier concentration; the Drude mobility $\mu = 1/e\rho n$ and the field-effect mobility $\mu_{FET} = gL/WC_{tot}V_{sd}$,[56] where $\rho$ stands for the resistivity, n – the carrier concentration, g – the transconductance, L – the channel length, W – the channel width, $C_{tot}$ – the total capacitance (see S5, Supplementary Data) and $V_{sd}$ – the source-drain voltage. A good agreement between the two models is shown in Figure 5 (b). The reported mobilities are consistent with values obtained from the Hall effect mobility measurements of solution-gated graphene on SiC.[38] It is worth to note that in our case these values should be treated as a lower bound for the estimation of the carrier mobilities, since we employed two point measurements, for which the contact resistances may significantly increase $\rho$ and lower $g$.[38] The access regions close to the contacts where the carrier concentration cannot be effectively steered by the applied gate voltage are other limiting factors, since they are covered with the passivation resist and thus can give rise to additional doping-dependent junctions. Indeed, a Hall effect measurement using the van der Pauw configuration of the bare sample (1cm x 1cm) prior to the processing yielded an n-type carrier concentration $n \approx 7.5 \cdot 10^{12}/cm^2$ and mobility $\mu \approx 860 \ cm^2/Vs$. The dependence of the mobility on carrier concentration can be approximated by a power law of $\mu \propto n^{-0.72}$, see red line in Figure 5 (b). The value of the exponent fits in the range of reported values for graphene on the Si face of SiC from almost -1 measured for as grown samples [57] to $\approx -0.3$ for a SGFET in an aqueous solution.[38]

### 3.4. Spectroelectrochemistry

The results presented so far in section 3.2 and 3.3 were based on measurements performed in a limited gate voltage window. For this voltage range, no electrochemical reactions occur, thus we could describe our observations within the notion of an ideal polarized electrode (see S4, Supplementary Data). In order to trigger electrochemical reactions, we intentionally increased the gate voltages. The experimental setup, as described in Figure 1, remained unchanged.

Figure 6 (a) presents a false color plot of the Raman spectrum obtained as a function of the gate voltage, which visualizes the impact of voltages beyond the notion of an ideal polarized electrode. The results were obtained using the same sample (non-intercalated) as in the previous Figures. The spectra were corrected for the SiC background (Supplementary Data). To visualize the effects more markedly, three cuts indicated as horizontal lines in Figure 6 (a) are shown in Figure 6 (b). At voltages around +0.6 V, one can observe the emergence of additional features in the Raman spectrum: a complex structure in the region around the D peak, a feature at 2100 cm$^{-1}$ and a broad band from

around 2830 cm$^{-1}$ to 2950 cm$^{-1}$. The peaks at ~1300 cm$^{-1}$, ~1440cm$^{-1}$, ~2855 cm$^{-1}$, ~2920 cm$^{-1}$ can be best assigned to the twisting, bending, in-phase and out-of-phase stretching vibrations of C-H bonds.[58-60] For the rather broad band at ~2100 cm$^{-1}$, we propose two possible assignments. Firstly, the broad character and the spectral position could be explained by the presence of sp$^1$ hybridized carbon.[61,62] Secondly, one could relate this feature to Si-H stretching vibrations.[60]

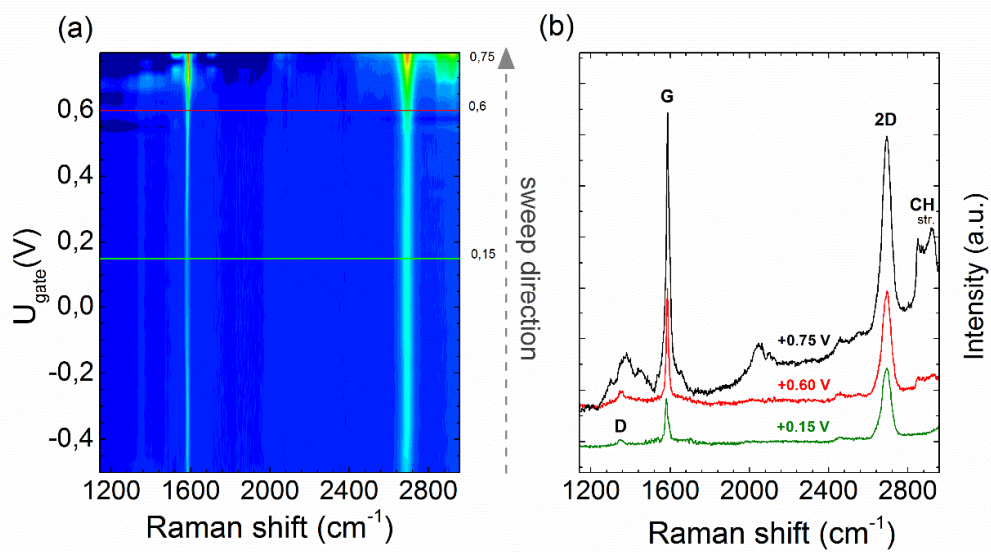

**Figure 6.** Raman spectra obtained using the same sample (no hydrogen intercalation) as in the previous Figures: (a) Raman false color contour plot; the sweep direction is from negative to positive gate voltages. The horizontal lines indicate the position of the spectra shown in (b). With increasing positive gate voltage, additional features at around 1300 cm$^{-1}$, 2100 cm$^{-1}$ and 2900 cm$^{-1}$ emerge and an increase in the D band intensity can be observed. Spectra in (b) are shifted vertically for clarity.

The results obtained in a similar manner for a hydrogen-intercalated sample are shown in Figure 7. A gate voltage sweep from negative to positive voltages is presented in Figure 7 (a). One can clearly observe a strong change in the Raman spectra for voltages exceeding +0.25 V. Three cuts indicated as horizontal lines in Figure 7 (a) are shown in Figure 7 (b). This change is reflected by an increase of the defect-related D band intensity (~ 1350 cm$^{-1}$) and the appearance of a sharp feature at around 2130 cm$^{-1}$ together with the bands that were already assigned to C-H vibrations as for the previous, non-intercalated sample. In contrast to Figure 6, the feature emerging at around 2130 cm$^{-1}$ is very sharp, therefore we can attribute it to Si-H stretching vibrations,[63] since the narrow shape can hardly be attributed to sp$^1$ hybridized carbon clusters. The Si-H stretching vibrations can regularly be observed in Attenuated Total Reflection infrared spectroscopy (ATR-FTIR) on hydrogen annealed SiC substrates and hydrogen-intercalated epitaxial graphene on SiC.[42,64,65] However, as a consequence of a small Raman cross section, it is difficult to observe these vibrations in Raman spectroscopy measurements in the visible range.[66, 67] We propose two different explanations for the emergence of this peak in our experiment. The first reason would be the formation of new Si-H bonds under the influence of the gate voltage. An alternative explanation would be the increase of the Raman cross-section of the modes corresponding to the Si-H bonds which already existed for hydrogen-intercalated graphene [42] before applying the gate voltage. The change of cross-section could result from structural changes caused by electrochemical reactions. The latter explanation is supported by a general rise in Raman intensity, as can be seen from an increase of the G and 2D bands upon the onset of reactions. Notably, the Raman intensity enhancement is valid only for the graphene-related peaks. The intensity of the 2$^{nd}$ order SiC background Raman signal did not show a similar enhancement. This effect can be clearly seen for the uncorrected data presented in the Supplementary Data (S2,S3). Since it has been reported that clusters of hydrogenated regions can form at the surface of graphene,[68] we speculate that plasmonic effects caused by these clusters could be the origin of the functionalized graphene Raman intensity enhancement. This explanation is in line with a recent report on the Raman signal enhancement on hydrogen-intercalated epitaxial graphene on SiC (0001) due to single microparticle mediated surface enhanced Raman spectroscopy (smSERS).[69] Using gold microparticles, the authors were able to observe a 13–20-fold enhancement of the G band intensity, which allowed the observation of the Si-H peak evolution as a result of hydrogen intercalation.

It is worth noting that the band at around 2900 cm$^{-1}$ can hardly be attributed to the D+D' band, since the position, the shape and the intensity ratio to the D peak do not match.[17] In fact, C-H features have been already observed in Raman electrochemical studies of epitaxial graphene on SiC in sulfuric acid,[70] for nanoporous carbons[35] and for dissociative $H_2$ adsorption on CVD and exfoliated graphene on $SiO_2$.[71] An increase of the luminescence background which we also observe in our experiments, was reported in ref 70 where it was ascribed to the hydrogenation of graphene. Such an increase in luminescence background has also been observed for increased hydrogen content in hydrogenated carbon films,[72] which further supports the hypothesis about the role of hydrogen in the observed reactions. Since the spectra in Figure 7 are corrected for the SiC background, this effect is not apparent. The effect becomes clearly observable on the uncorrected spectra, which can be found in the Supplementary Data (S2,S3). Figure 7 (c) shows results of a gate voltage sweep performed in opposite direction for the same sample. In between the measurements, the gate voltage was switched off for several hours and one can observe that the peaks vanish, which is in agreement with the reports on nanoporous carbon and points towards the reversibility of the reactions.[35] Upon application of a positive voltage beyond the threshold for the reactions, the peaks reappear, but not instantly. This indicates that the reactions are kinetically very slow in our system, since the presented results were recorded over a timespan of several hours. At a voltage of about +0.18 V (Figure 7 (d)), we can clearly identify the same features as in Figure 7 (b). Once again, the peaks vanish as a result of further reduction of the gate voltage, illustrating the reversibility of the process.

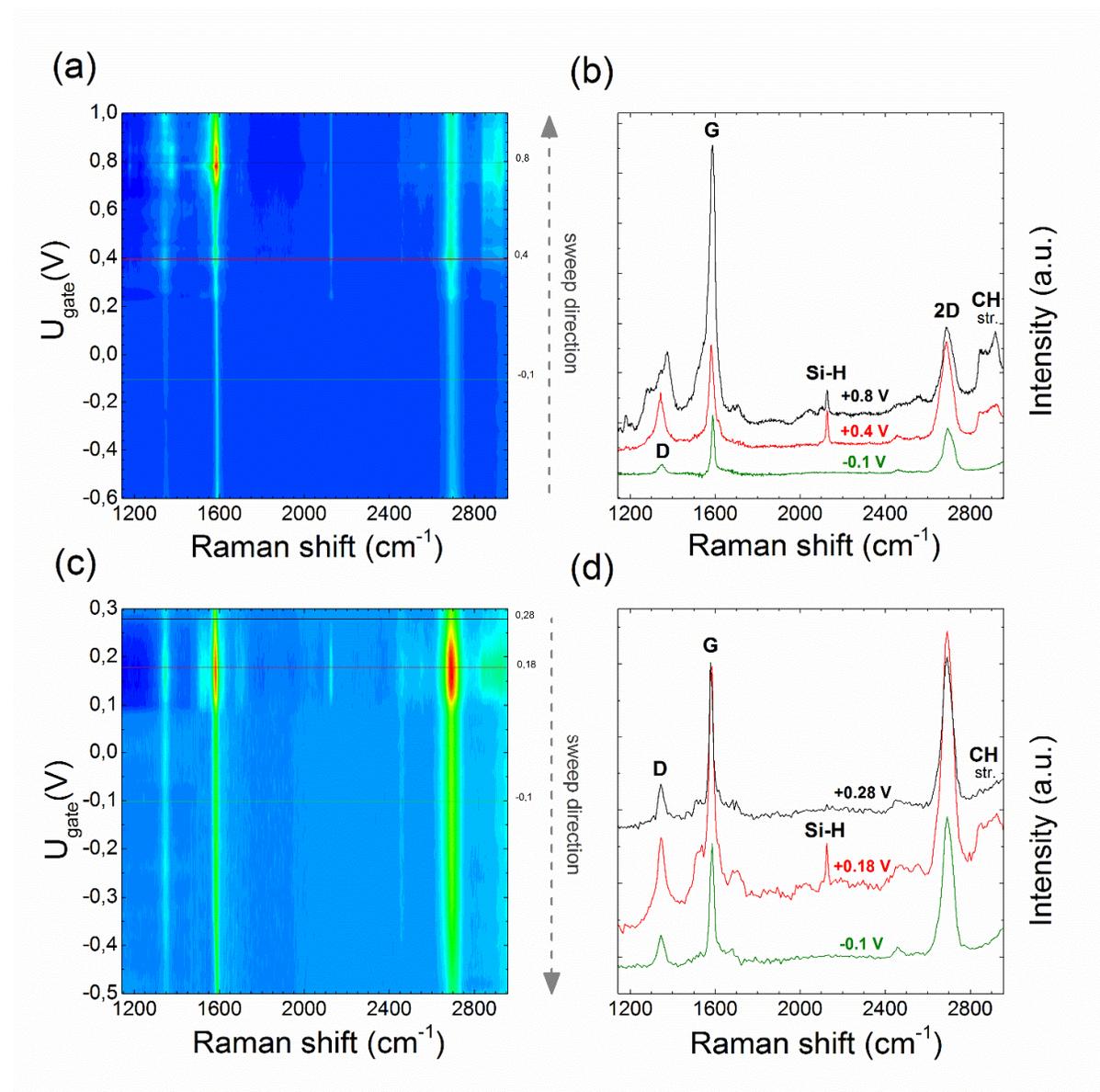

**Figure 7.** Raman spectra of a hydrogen-intercalated sample: (a) Raman false color contour plot; the sweep direction is from negative to positive gate voltages. The horizontal lines indicate the position of the spectra shown in (b). With increasing positive gate voltage, additional peaks (Si-H, C-H) emerge and an increase in the D band intensity can be observed. (c) Raman false color contour plot of a sweep in the opposite direction as compared to (a); the spectra corresponding to the horizontal lines are presented in (d). Similarly to (a), Si-H and C-H peaks emerge together with an increased D band. A further decrease in the gate voltage, however, leads to a decrease in the D band and the disappearance of the Si-H and C-H peaks indicating that the processes are reversible. Spectra in (b) and (d) are shifted vertically for clarity.

A comparison of the two green spectra for $U_{gate}$= -0.1 V plotted in Figures 7 (b) and (d), however, shows that despite the reduction in the D peak, it does not decrease to the same level as before the voltage sweeps. This points towards a partial reversibility.[73] In principle, one cannot exclude the impact of a prolonged exposure of the sample to the laser beam. However, the laser exposure cannot explain the appearance and disappearance of the hydrogen-related peaks upon the change of the gate voltage for different sweep directions. Furthermore, we observe a good agreement between the onset of current passing through the solution gate (see S6, Supplementary Data), which is an indication for the onset of an electrochemical reaction, and the appearance of the Raman features. To further prove that the influence of a prolonged laser exposure is not responsible for the observed D band increase, we present the D band region of the Raman spectra corresponding to Figure 6 and Figure 7 in the Supplementary Data (S7, S8). No significant change of the D band for different gate values can be observed below the reaction threshold voltage.

The behavior of the Raman D band in this work is in line with a recent report on Raman measurements of gated defective graphene samples on $SiO_2$.[74] An increase of the D band intensity for higher gate voltages was reported and attributed to electrochemical reactions occurring in the presence of water trapped at the $SiO_2$/graphene interface. Studies on gated graphene FET on $SiO_2$ in different atmospheres[75,76] showed that the hysteresis commonly observed under ambient conditions can also be attributed to electrochemical doping caused by water at the graphene/$SiO_2$ interface. It has been further shown that the number of graphene layers is essential for the determination of graphene's chemical behavior.[77-79] A higher reactivity can be expected for monolayer than for bilayer graphene. Due to the terraced structure of the SiC substrate, the actual morphology of graphene on the surface of our sample is much more complex[29,52,80] than for exfoliated graphene. The difference of the electrochemical behavior on the terraces and at the step edges is, however, beyond the scope of this paper.

We ascribe the appearance of the observed Raman features to partially reversible processes occurring after the chemisorption of hydrogen under cathodic conditions on the graphene working electrode.[35,70] The actual experimental mechanism, however, is still not fully explored, which stresses the importance of further in-depth spectroelectrochemical studies.

## 4. Conclusions

In conclusion, we performed novel combined electro-optical measurements of aqueous solution-gated graphene FETs on 4H-SiC. For epitaxial bilayer graphene on SiC we were able to observe the resonant electron-phonon coupling effect at room temperature. Based on this we deduced the geometrical capacitance, thus providing a direct calibration of the Fermi level versus gate voltage for our system. This allowed us to draw conclusions about the dependence of the carrier mobility on carrier concentration for the transistor. By intentionally applying high gate voltages, we were able to induce the chemisorption of hydrogen on our bilayer graphene sample. We traced the chemisorption process by in-situ Raman spectroscopy, observing the appearance of Si-H and C-H modes and an increase of the D band intensity. The process was partially reversible upon application of negative gate voltage values. Hence, by changing the applied gate voltages, we obtained an electrical switch to the chemisorption of hydrogen on graphene. The presented results highlight the great potential of the experimental setup. Our approach gives insights into the electron-phonon coupling in bilayer graphene on SiC. At the same time it allows for spectroelectrochemical measurements on the micrometer scale, which may help to shed more light on sorption processes on graphene.

**Acknowledgements**

This work was supported by the Foundation for Polish Science International PhD Projects Program co-financed by the EU European Regional Development Fund, the National Centre for Research and Development project GRAF-TECH/NCBR/02/19/2012, the National Science Centre Poland grants 2014/13/N/ST3/03772, 2013/10/M/ST3/00791, 2012/07/B/ST3/03220 and the European Union Seventh Framework Program under grant agreement n°604391Graphene Flagship.

**Supplementary data**

Raman background subtraction procedure, 2D band shape, uncorrected spectral maps, cyclic voltammetry, total capacitance, current onsets and D band evolution.

Supplementary Data for

# In-situ Raman spectroscopy of the graphene/water interface of a solution-gated field effect transistor: electron-phonon coupling and spectroelectrochemistry

J. Binder[1]*, J. M. Urban[1], R. Stepniewski[1], W. Strupinski[2] and A. Wysmolek[1]

[1] Faculty of Physics, University of Warsaw, Pasteura 5, 02-093 Warsaw, Poland
[2] Institute of Electronic Materials Technology, Wólczyńska 133, 01-919 Warsaw, Poland

1. **SiC Raman background subtraction procedure**

In order to be able to observe the Raman features related to graphene, we employed the following method. After focusing on the graphene surface, the automated z-table was moved up by 30 μm so that the laser was focused inside the SiC substrate. Thanks to our confocal setup, the spectrum recorded in such a way contained predominately information originating from the SiC substrate. We thus obtained our SiC reference spectrum. An automated script was used to perform the background subtraction. A linear function was fitted and subtracted from the raw data in the region from around 2000 cm$^{-1}$ to 2400 cm$^{-1}$ for both the reference and each of the measured graphene spectra. This allowed to eliminate the differences in the luminescence background. An integration of the intensity was performed in the region around 1650 cm$^{-1}$ to 1800 cm$^{-1}$ for the SiC reference spectra as well as for each graphene spectrum. The ratio of these two integrated values was used as the scaling factor for the subtraction of the reference SiC signal in order to account for intensity differences.

With this method, we obtained a reliable algorithm that not only eliminates the SiC background, but also provides information about the luminescence background (slope of the linear fit) and the SiC background intensity (scaling factor for background subtraction).

## 2. 2D band shape

The shape of the 2D band shown in S1 is typical for AB stacked bilayer graphene. Four Lorentzians with a width of 24 cm$^{-1}$ were used to fit the spectrum.[1]

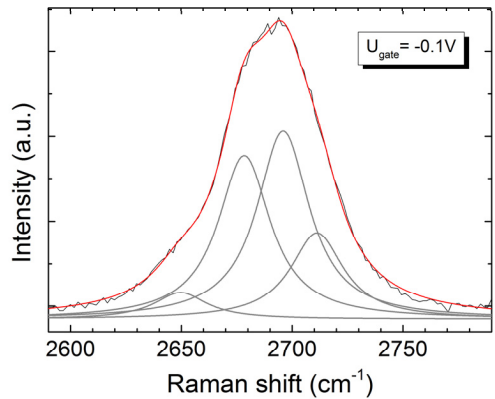

S1: Raman spectra in the 2D region for an applied gate voltage of -0.1V.

## 3. Uncorrected spectral maps

S2 shows the raw data corresponding to Fig. 6 in the main article. A clear increase in Raman intensity can be observed for voltages higher than +0.5 V. The spectra in S2 (b) are the uncorrected spectra (no vertical shifting). One can observe that for higher voltages the signal corresponding to graphene increases (G and 2D band), whereas the 2$^{nd}$ order SiC background signal does not show such an increase, indicating that an enhancement mechanism is only present for graphene.

A rise in the luminescence background manifested by a linear increase towards higher energies can be recognized in S2 (b). This is in agreement with reports[2,3] in which an increase of luminescence background upon increase of the hydrogen content of the carbon was observed.

S3 shows the raw data of the sweep presented in Fig. 7 in the main article. Similar to S2, we observe an intensity enhancement of the graphene-related features and an increase in the luminescence background.

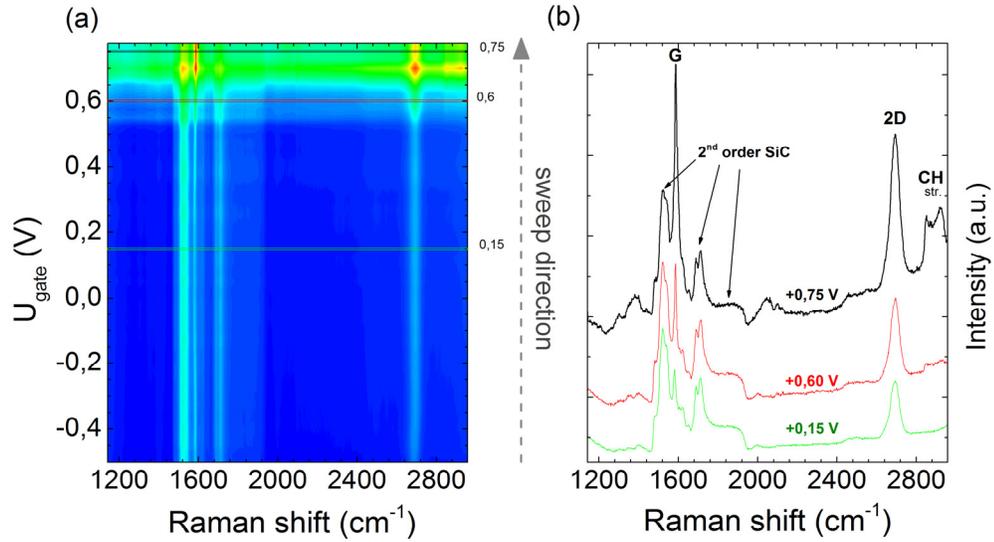

S2: Raw data of the non-intercalated sample (compare to Fig.6): (a) Raman false color contour plot; the sweep direction is from negative to positive gate voltages. The horizontal lines indicate the position of the spectra shown in (b). With increasing positive gate voltage, additional features at around 2100 cm$^{-1}$ and 2900 cm$^{-1}$ emerge and an increase in the D band intensity can be observed. Spectra in (b) correspond to the lines in (a). The spectra are not vertically shifted and the increase in luminescence as well a general increase of the G and 2 D band can be observed.

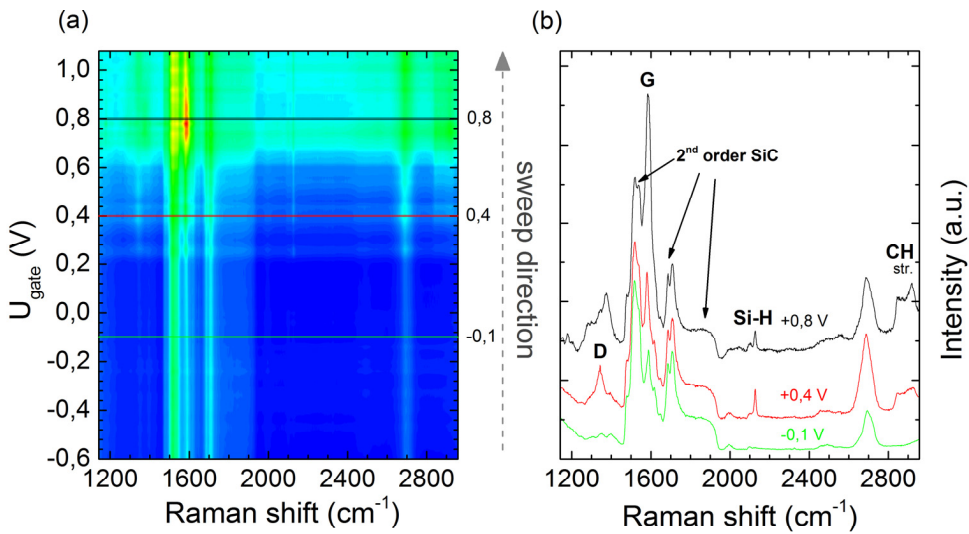

S3: Raw data of the hydrogen-intercalated sample (compare to Fig.7): (a) Raman false color contour plot; the sweep direction is from negative to positive gate voltages. The horizontal lines indicate the position of the spectra shown in (b). With increasing positive gate voltage additional features at around 2100 cm$^{-1}$ and 2900 cm$^{-1}$ emerge and an increase in the D band intensity can be observed. Spectra in (b) correspond to the lines in (a). The spectra are not vertically shifted and the increase in luminescence as well a general increase of the G and 2 D band can be observed.

## 4. Cyclic Voltammetry

In order to verify whether it is adequate to assume a constant geometrical capacitance, we performed cyclic voltammetry (CV) measurements using our sample mount and the configuration as presented in Figure 1 in the main text. The results for different scan rates are shown in S4. Even at high scan rates (up to 800 mV/s), the CV graphs do not show any specific peaks in current that could be related to redox reactions. Therefore, we employ the notion of an ideal polarized electrode. The curves can be described by taking two components into account. The first component is the current due to capacitive charging of the electrical double layer, which leads to a signature in the characteristics that is similar to the charging behavior of a regular capacitor. The second linear component is due to the background current caused by impurities in the solution.[4] It has to be mentioned that because of the closed reservoir and the small dimension of the sample fixture no purging gas like argon or nitrogen was used to remove oxygen from the solution. By implementing the possibility to purge the electrolyte, the contribution of the background current to the measurements is expected to decrease.

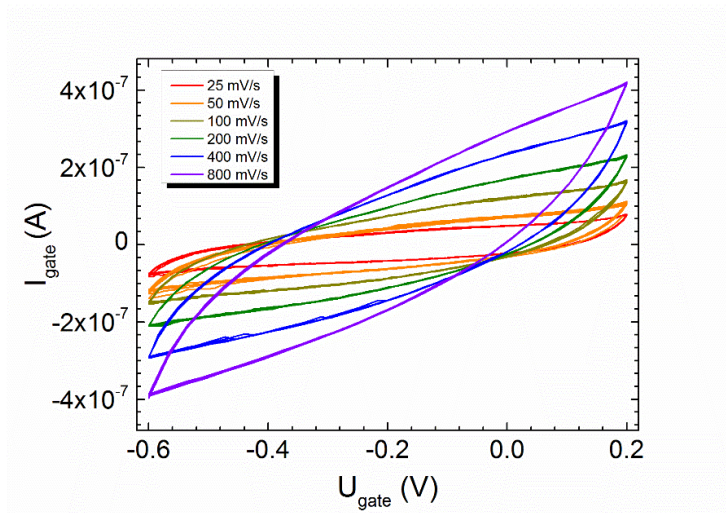

S4: Cyclic Voltammetry measurements with different scan rates performed using the experimental setup shown in Figure 1 in the main text.

## 5. Total Capacitance

The results of the capacitance calculations are presented in S5. By employing a gate voltage difference of 0.325 V between the minima observed in Raman spectroscopy (Fig. 3 (b)), we extracted a geometrical capacitance of 6.65 $\mu F/cm^2$ (red dashed line), which we assumed to be constant, also based on our CV results. The blue dash-dotted line shows the calculated quantum capacitance for bilayer graphene based on Ando's calculations[5] and Luryi's[6] definition of quantum capacitance. The black solid line shows the result of both capacitors connected in series giving the effective capacitance of the system.

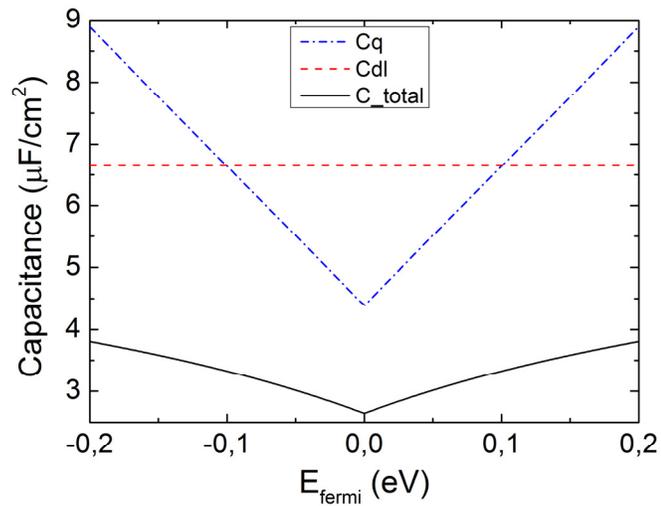

S5: Results of the capacitance calculations.

## 6. Current onset

As mentioned in the main text, the current onsets coincided with the appearance of the Si-H and C-H Raman features in the spectra. The shift that can be observed between the forward and backward sweep is also apparent in the current.

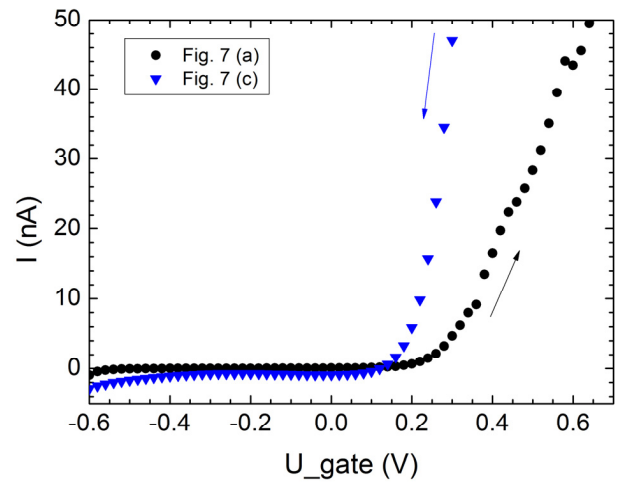

S6: Onset of currents for Raman maps presented in Fig.7 (a) and (c)

## 7. D peak evolution

S7 presents the spectra for $U_{gate}$= -0.4 V, 0 V, +0.4 V in the D and G band region, corresponding to Fig.6 or S2. The spectra are background corrected, but not normalized. It can be seen that the D band, even after prolonged exposure to laser irradiation, does not increase. This stresses the fact that the Raman features emerging at higher voltages do not result from the laser exposure. The same is true for the hydrogen-passivated sample. No significant increase of the D band can be observed upon prolonged laser exposure, as manifested in S8.

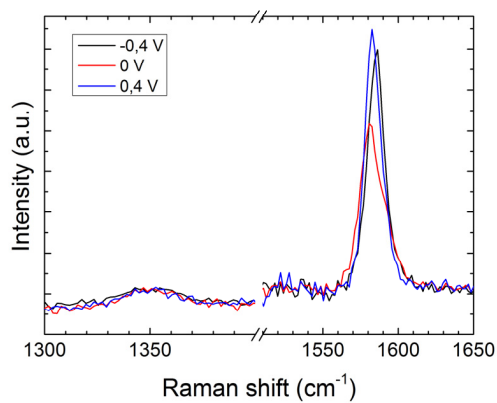

S7: Raman D and G band region of the non-intercalated sample. The spectra were background corrected, but were not normalized. (compare Fig.6, S2)

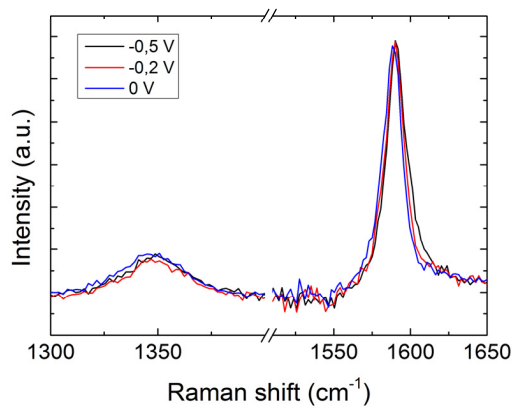

S8: Raman D and G band region of the intercalated sample. The spectra were background corrected, but were not normalized. (compare Fig. 7, S3)